\documentclass[a4paper,fleqn,usenatbib]{mnras}
\usepackage[T1]{fontenc}
\usepackage{ae,aecompl}
\usepackage{graphicx} % Including figure files
\usepackage{amsmath} % Advanced maths commands
\usepackage{amssymb} % Extra maths symbols
\title[A new 2.2 Mpc Giant Radio Galaxy at z=0.57]{Discovery of a new, 2.2 Mpc Giant Radio Galaxy at a redshift of 0.57}
\author[Sebastian, Biny et. al.]{Biny Sebastian$^{1}$, C.H. Ishwara-Chandra$^{1}$, Ravi Joshi$^{2,3}$, Yogesh Wadadekar$^{1}$,\thanks{E-mail: biny@ncra.tifr.res.in (BS)}
\\
$^{1}$National Centre for Radio Astrophysics, TIFR, Post Bag No. 3, Ganeshkhind Post, Pune, India\\
$^{2}$Inter University Centre for Astronomy and Astrophysics, Post Bag No. 4, Ganeshkhind Post, Pune, India\\
$^{3}$Kavli Institute for Astronomy and Astrophysics, Peking University, Beijing 100871, China
\\
}
\date{Accepted XXX. Received YYY; in original form ZZZ}
\pubyear{2017}
\begin{document}
\label{firstpage}
\pagerange{\pageref{firstpage}--\pageref{lastpage}}
\maketitle

% Abstract of the paper
\begin{abstract}
We report the discovery of one of the largest and most distant Giant Radio Galaxy (GRG)
in the Lynx field which was discovered using deep Giant Metre-wave Radio Telescope (GMRT) 150 MHz observations. 
 The core is detected at 150 MHz and also in the VLA FIRST survey.
Spectroscopic observations carried out using the IUCAA Giravali Observatory(IGO) provided a  redshift value of 0.57. This redshift was later confirmed with data from the Sloan Digital Sky Survey (Data Release 12). The angular size of the GRG is 5.5 arcmin and at the 
redshift of 0.57, its linear size is 2.2 Mpc.
At this high redshift, only a few radio sources are known to have such
large linear size. In order to estimate the spectral index of the bridge emission as well as the spectral age of the source, we observed this source at L-band, 610 MHz and 325 MHz bands with the GMRT. We present the spectral ageing analysis of the source which puts an upper limit of 20 Myr on the spectral age. The better resolution maps presented here as opposed to the original 150 MHz map shows evidence for a second episode of emission. We also find that the core is detected at all four frequencies with a spectral index of 0.85, which is steeper than normal, hence we speculate that the core may be a compact steep spectrum source (CSS), which makes this giant radio galaxy a candidate triple-double radio galaxy.

\end{abstract}
% Select between one and six entries from the list of approved keywords.
% Don't make up new ones.

%%%%%%%%%%%%%%%%% BODY OF PAPER %%%%%%%%%%%%%%%%%%

\section{Introduction}
A small fraction of active galaxies (AGNs) exhibits powerful radio emission on either side of the nuclei of the galaxies. The majority of such radio sources are compact in size. However, a tiny fraction of them do attain sizes over a mega parsec and are known as 'Giant Radio Sources' (GRS). Their large linear size ($>$ 1 Mpc) makes them interesting candidates to understand the evolution of radio sources and to study the ambient intergalactic medium which confines the lobes so far away from the parent galaxy. However, detecting such giant radio sources becomes difficult because the bridge emission connecting the two lobes is often not seen. Low-frequency radio surveys are better suited to search for this population \citep{2000A&AS..146..293S} as compared to higher frequency radio surveys because the aged plasma will have a steep spectrum and is brighter at low frequencies.% {\it YW: perhaps mention here why it is easier to detect lobes at low frequency} 

Till date, more than 300 giant radio galaxies have been discovered (\cite{2017arXiv170400516D}; \cite{1999MNRAS.309..100I}; \cite{2001A&A...374..861S}; \cite{2001A&A...370..409L}, \cite{1996MNRAS.279..257S}; \cite{2001A&A...371..445M}, \cite{2006A&A...454...85M}, \cite{2005AJ....130..896S}, \cite{2014AstBu..69..141S}), \cite{2015MNRAS.453.2438T}). The longest identified giant radio galaxy is J1420-0545 with a linear size of 4.69 Mpc (\cite{2008ApJ...679..149M}). \cite{2016ApJS..224...18P} classified $~$1700 radio sources from the NVSS catalog as giant radio sources based on their angular extent using a pattern recognition algorithm. Because large angular extents are used as primary criteria while classifying giant radio galaxies, the number of giant radio sources identified at high redshifts is low. Most of them are nearby ($z  \lesssim 0.5$) and with a linear size smaller than 2 Mpc. Increasing the number of identified giant radio galaxies at high redshift is necessary for studying the evolution of the radio galaxy population over redshift.

Numerous studies were undertaken to understand the reason for the large linear sizes of these sources. 
The dynamical age of these sources suggest these have evolved over long periods of time \citep{1997MNRAS.286..215K}. However, \cite{1998A&A...329..431M} find spectral ages comparable to the radio galaxies with normal sizes. Hence, they conclude the large linear sizes are probably due to the low density of the inter-galactic medium surrounding the giant radio sources. Many authors have also investigated the role of the nuclear host power in creating these giant sources. \cite{1999MNRAS.309..100I} find that the giant radio galaxies have similar core strengths compared to the rest of the radio galaxies of normal sizes having comparable luminosity.  
It is also interesting to note that several of the giant radio sources show recurrent jet activity (\cite{2013MNRAS.436.1595K}, Kronberg et al, 2001). Though the mechanism which leads to the stopping and restarting of radio galaxies is still not well understood, there are models which invoke galaxy mergers to explain the episodic jet activity. For example, \cite{2003MNRAS.340..411L} suggests that the disruption of the accretion disk due to the inspiral of a second SMBH into the larger one causes an interruption in the jet activity. 

The spectral ageing analysis is an invaluable tool in tracing the evolution of radio galaxies. Due to the large angular extent of giant radio galaxies, we can learn more about various energy loss and gain mechanisms of the radiating particles at different locations individually like the hotspots, lobes, etc.  It was noted previously by \cite{1999MNRAS.309..100I} and Schoenmakers et al. 2000a that inverse Compton losses dominate the synchrotron losses in the lobes of giant radio galaxies. Moreover, galaxies at higher redshifts are expected to have higher inverse Compton losses due to denser cosmic microwave background. Also understanding spectral index variations across the lobes in a normal population of radio sources at higher redshifts is hard because of their low angular widths. Hence giant radio galaxies at higher redshifts help us get a better handle on the spectral ageing and evolution studies at higher redshifts.

In this paper, we report the discovery of a giant radio source of linear
size 2.2 Mpc at 0.57 redshift in the LBDS-Lynx field. The paper is organised as follows. The details of observations and data analysis techniques used are elaborated in Section~\ref{obs}. Section~\ref{res} discusses several results derived from morphological and spectral ageing analysis. 

We used a cosmology, with $\Omega_M$ =0.286, $\Omega_{baryon}$ = 0.0463 and $H_0$ =69.3 km s$^{-1}$ Mpc$^{-1}$.
%{\it Specify WHICH COSMOLOGY USED FOR LINEAR SIZE.}
%Evolution, confinement, possibility of multiple epoch

\section{Observations and Data Analysis}
\label{obs}
The LBDS(Leiden Berkeley Deep Survey)-Lynx field was observed with Giant-Metrewave Radio Telescope (GMRT) at 150 MHz to search for high-redshift radio galaxies \citep{2010MNRAS.405..436I}. A candidate giant radio source was found in the field, beyond the half-power beam width with an RA DEC of 08h44m08.8s +46d27m44s.% {\it should give the RA/Dec of the radio galaxy here. Maybe a table with VLA position, GMRT position, and SDSS position can be added.} 

The core was also detected in the VLA FIRST survey data. An SDSS counterpart was identified with a red galaxy having a photometric redshift of 0.62 and R band magnitude of 21.9 \citep{2009ApJS..182..543A}.  The angular size of the radio source was 5.5$'$ and at a redshift of 0.62, the linear size of the galaxy is $\sim$ 2.3 Mpc.
In the 150 MHz image, the radio lobes were barely seen. To further
characterize the giant radio galaxy, deep radio observations with GMRT
and spectroscopic observations with IUCAA Girawali
2.0 m optical telescope were carried out. The details of the radio and optical observations are given below.

\subsection{GMRT observations and data analysis}
GMRT observations were carried out at three frequencies, 325 MHz (P-band),
610 MHz, 1250 MHz (L-band). The details of the GMRT observations used
for imaging are given in Table~\ref{Observation details}.
The observations at 610 MHz and 325 MHz were carried out using the legacy system.
The observation in L-band was done using the new upgraded GMRT (uGMRT) with a wideband backend mostly using central square antennas to detect the bridge emission. The basic editing was carried out using the Astronomical Imaging Processing Software (AIPS). Bad data were identified and
flagged, and calibration was carried out using standard tasks. Radio Frequency
Interference (RFI) was identified and flagged after bandpass calibration. The bandpass calibration was rerun after the flagging of RFI. The
target file was SPLIT after final calibration. For data sets at 325 MHz and
610 MHz, images were made in AIPS using standard procedures. Several rounds of phase-only self-calibration were done to remove phase errors due to the ionosphere. Since the L-band data had a bandwidth of 400 MHz,
it was not possible to image in AIPS due to the unavailability of the MS-MFS
(Multi-Scale- Multi Frequency Synthesis) algorithm. So the imaging
was done in CASA using the task 'clean.' There were two data sets in L-band, which were
combined in UV plane using the task CONCAT. 
The wide band data were also divided into 8
split files of 50 MHz bandwidth each. Each split file was then imaged and self-calibrated individually after which all of them were combined back
in UV-plane in CASA for final imaging. 
As a final step, primary beam correction was done for all the images.
The angular size of the source was $\sim$ 5 arcmin  which is well within the HPBW (28 arcmins),
therefore single primary beam correction at L-band at the central frequency of the band
is justified.

For the spectral ageing analysis, images were made with matching uvranges at different frequencies. Uniform weighting was used while making these images. This procedure is adopted to make sure that the amount of diffuse emission picked up at different frequencies is similar.  The relative shift between the two maps was corrected for and was then convolved to the same beam before making the spectral index maps. Spectral index maps were made from 610 and 325 MHz maps using task COMB in AIPS. 

\begin{figure}
 % To include a figure from a file named example.*
 % Allowable file formats are eps or ps if compiling using latex
 % or pdf, png, jpg if compiling using pdflatex
 \includegraphics[width=\columnwidth]{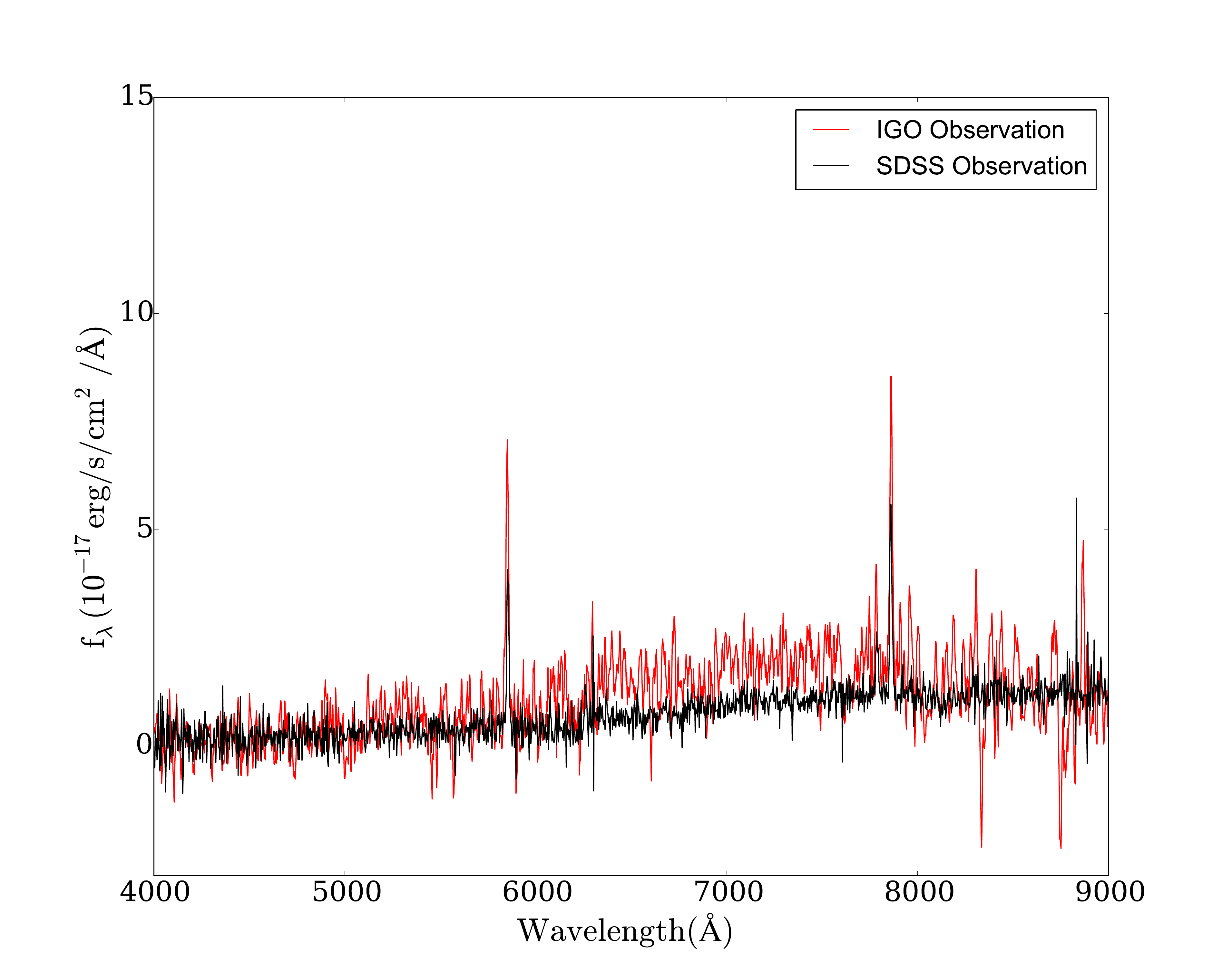}
    \caption{Optical Spectrum of J084408+462744: The spectrum in red was obtained using observations from the 2-m telescope at IUCAA Girawali Observatory. IGO spectrum is not flux calibrated. The spectrum in black was obtained from SDSS.}
    \label{IGO spectrum}
\end{figure}

\begin{table*}
 \centering
 \caption{GMRT Observation Details }
 \label{Observation details} 
 \begin{tabular}{ p{2.0cm} p{2.0cm} p{2.0cm} p{1.5cm}p{1.8cm}p{3.0cm}p{2.4cm}}
\\
\hline
Central Frequency (MHz)& Flux Calibrator&Phase Calibrator&Bandwidth& Time on source & Number of antennas &Observation Date\\ 
\hline
325  & 3C147 &0834+555 &32MHz& 3.7h &27 &15-Sep-2009\\
610  &3C147, 3C286 &0834+555 &32 MHz &8h &29 &06-May-2010\\
 &3C147, 3C286 &0834+555 &32 MHz &3h &29 &29-Nov-2015\\

1250  &3C147 &0834+555 &400MHz &2.5h &16& 01-Sep-2015\\
& & & & & (11/16 Central Square) &\\
 &3C147,3C286& 0834+555& 400MHz& 5h& 16 &07-Nov-2015\\
 & & & & & (14/16 Central Square) &\\
\hline

  \end{tabular}
\end{table*}

\subsection{Optical Spectroscopic Observations}

Spectroscopic observations were carried out using the 2 m optical
telescope at the IUCAA Girawali Observatory (IGO)  to find out the
spectroscopic redshift.  O[II] and O[III] lines are clearly detected.
The spectrum is shown in Figure~\ref{IGO spectrum}. The spectroscopic redshift
was estimated to be 0.5692$\pm$0.0001. Subsequently, independent spectroscopic observations from the SDSS DR12 data release \citep{2015ApJS..219...12A} of the corresponding object, SDSS J084408.85$+$462744.2 confirmed the redshift  %{\it which data release? Cite that DR paper} which turned out to be 0.56928$\pm$0.00006. 
%{\it YW: What is the objid of this galaxy? Are there any redshift flags? Any other notable features of the optical spectrum?}

%\subsection{Data Analysis}

\section{Results}
\label{res}
Here we present the results of radio observations and spectral ageing analysis.
The core and bridge emission is clearly detected at 325, 610 and 1400 MHz.
The core is unresolved in all the three bands. The details are as below.

\subsection{Morphology}
\begin{figure}
\begin{center}
\includegraphics[width=8cm]{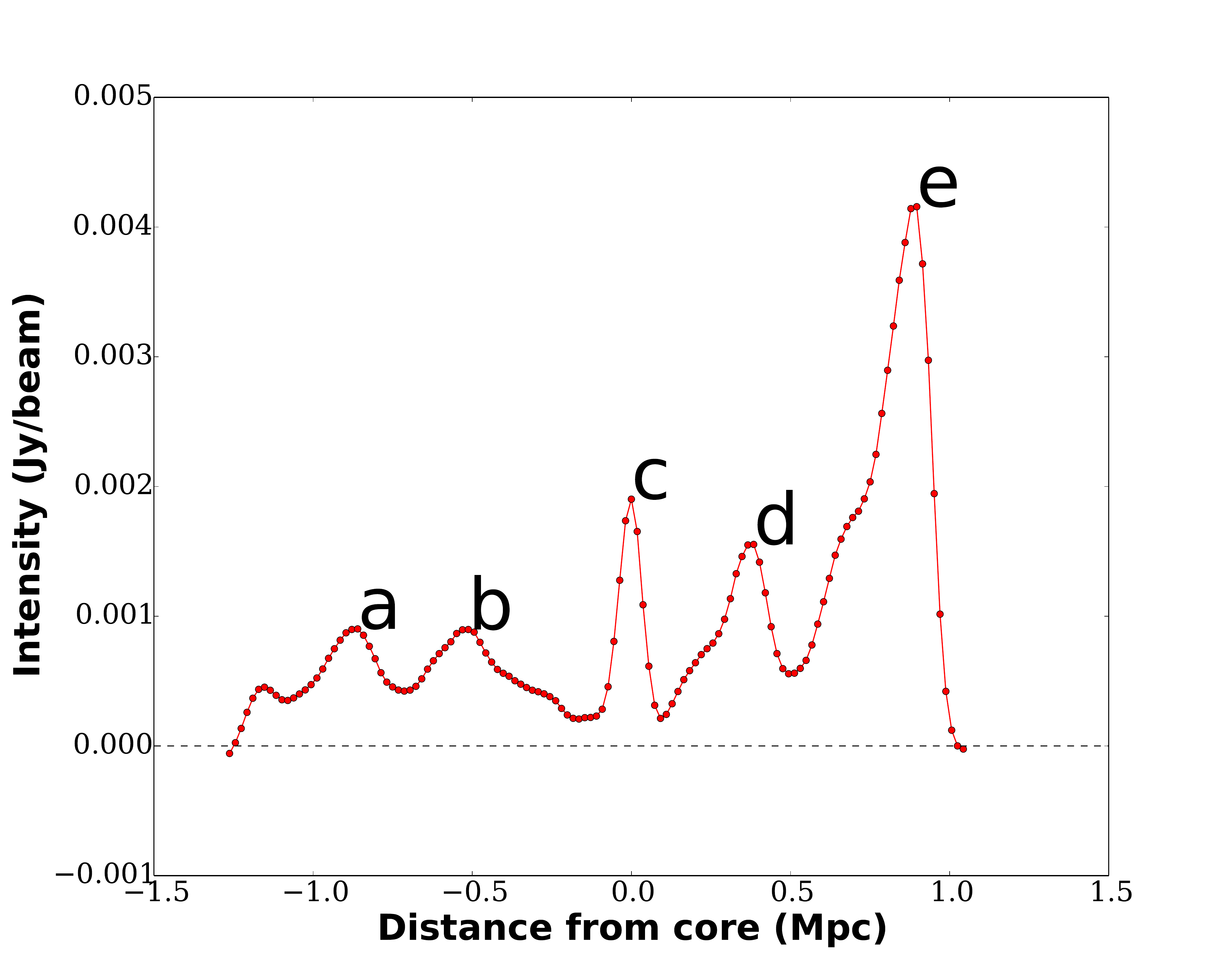}
\\

\caption[Different regions taken to calculate spectral age]
{The spatial profile of intensity along the length of the giant radio galaxy}
\label{spatialprofile}
\end{center}
\end{figure}

\begin{figure*}
\begin{center}
\begin{tabular}{cc}
\includegraphics[width=5.5cm,height=5.5cm]{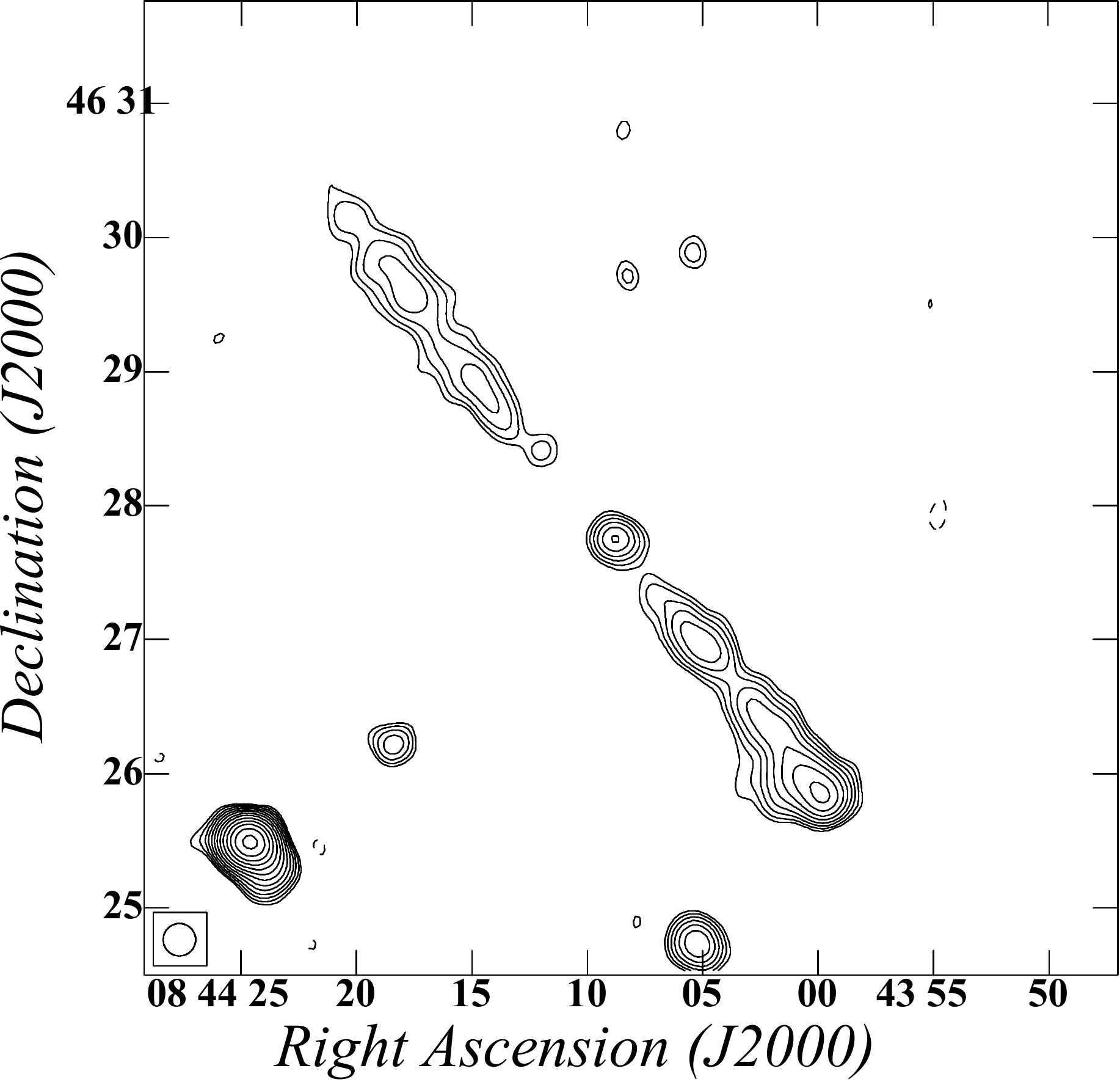} 
\includegraphics[width=5.5cm,height=5.5cm]{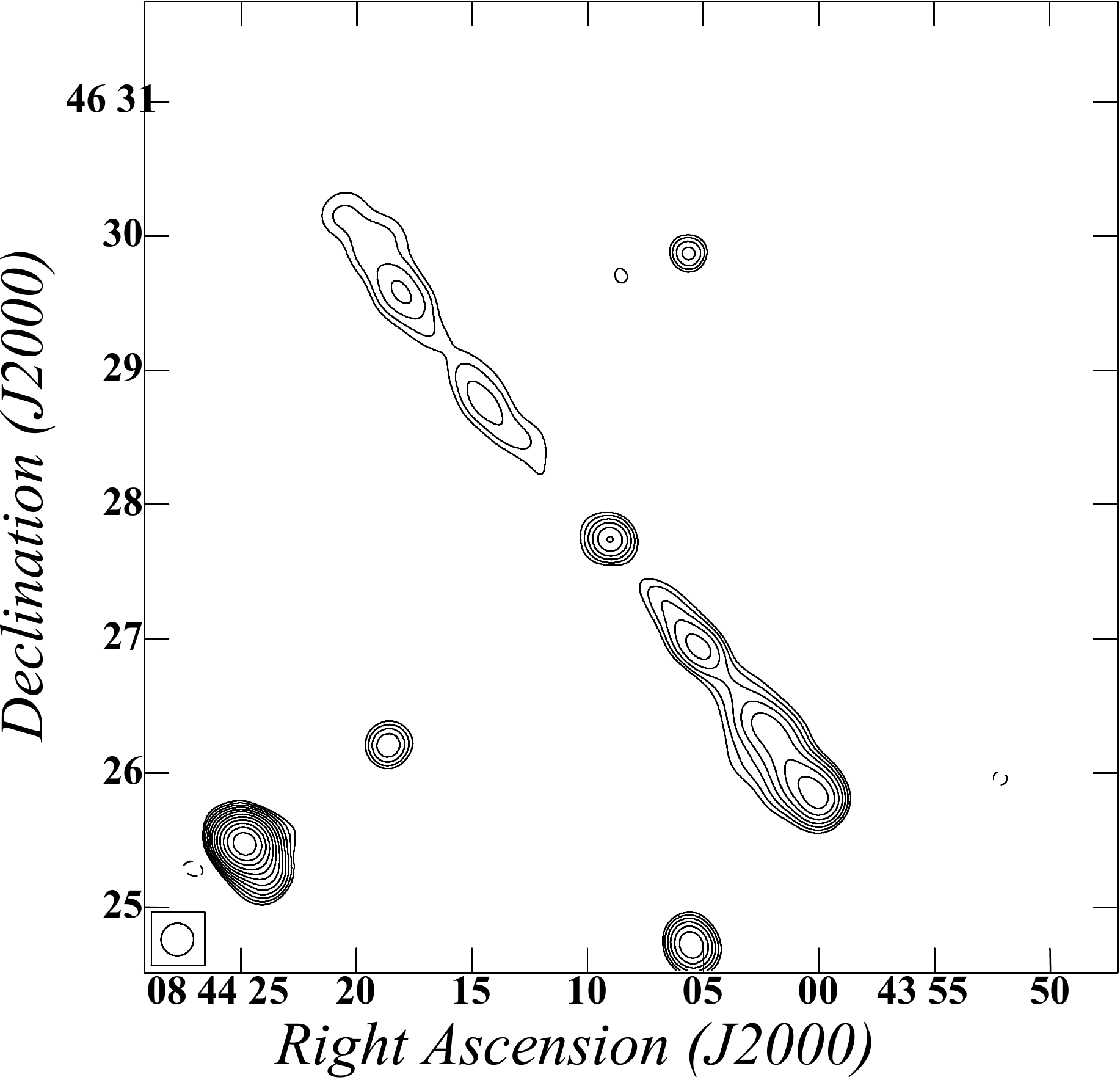} 
\includegraphics[width=5.5cm,height=5.5cm]{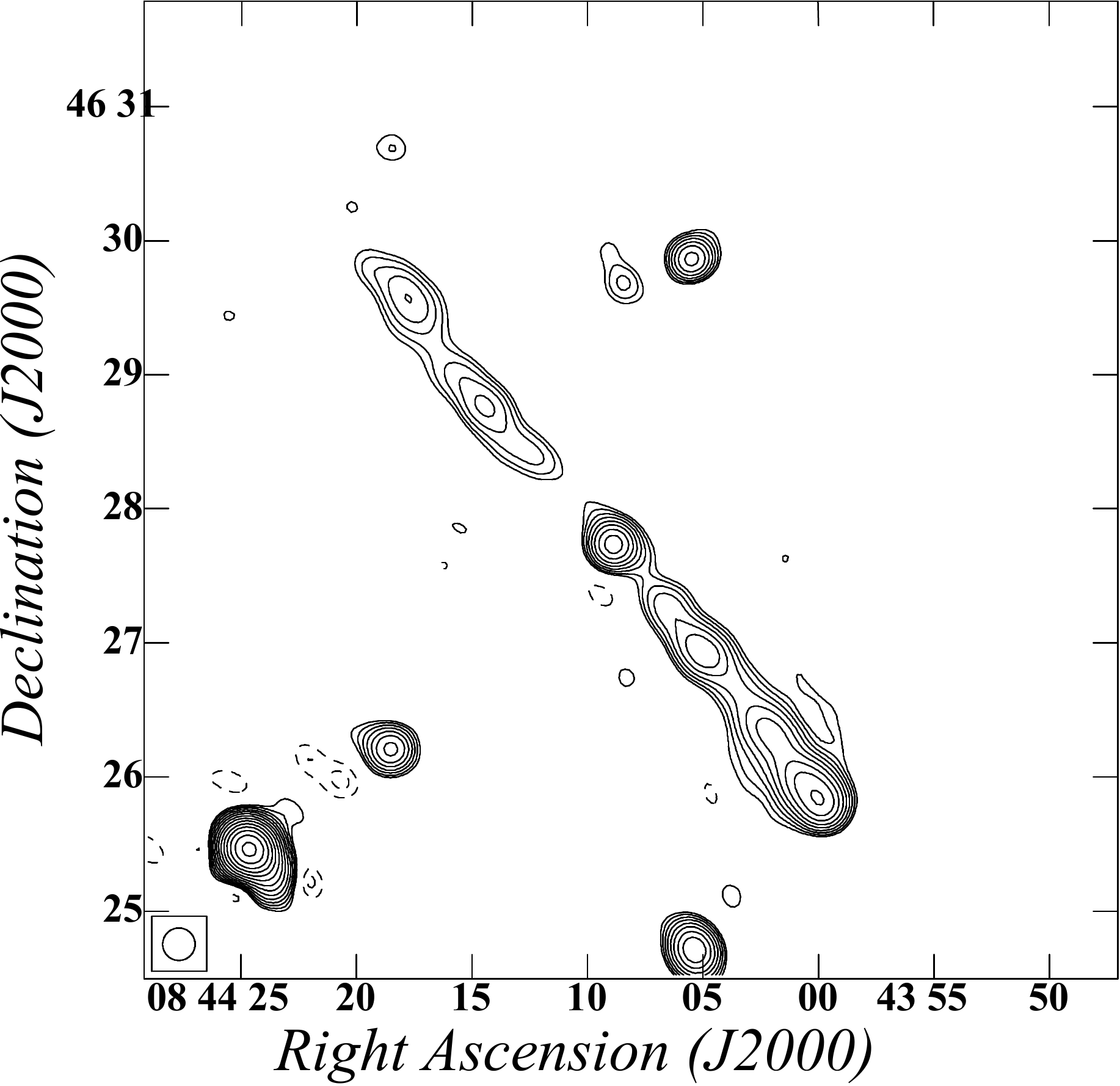}  
\end{tabular}

 \caption{Left-hand panel: Image at 325 MHz, peak surface brightness is  57.2 mJy~beam$^{-1}$, the rms is 0.2 mJy~beam$^{-1}$ Middle panel: Image at 610 MHz, the peak surface brightness is 25.5 mJy~beam$^{-1}$, the rms is 0.1 mJy~beam$^{-1}$ Right hand panel: Image at 1250 MHz, the peak surface brightness is 12.4 mJy~beam$^{-1}$, the rms is 0.03 mJy~beam$^{-1}$. Contour levels of all maps are 4 $\sigma \times$(-2, -1.40, -1, 1, 1.400, 2, 2.800, 4, 5.600, 8, 11.20, 16, 23). All maps have been convolved to a beam size of 14$^{\prime\prime}$.5 $\times$ 14$^{\prime\prime}$.5}
\label{what}
\end{center}
\end{figure*}

The contour maps at 4$\sigma$ level for all the three frequencies are shown in Figure~\ref{what}. The giant radio galaxy has a linear morphology with no noticeable distortions. \cite{2009ApJ...695..156S} points out that giant radio galaxies, in general, are free from any winged emission or distortions in the bridges. They also show that the jet axis of giant radio galaxy population is commonly aligned along the optical minor axis. Hence, the absence of distortions suggests that the large sizes of this giant galaxy might also be due to the ejection in a direction of least resistance or the steepest pressure gradient. The core is bright indicating the presence of ongoing
activity. The hotspots, core and the lobes are clearly detected in all
three frequency maps. The jet is not detected. A closer look at the lobes towards south shows tightening of contours almost halfway between the
core and the final hotspot. A similar brightening in flux density is
found midway on the other side as well. If this tightening is arising from yet another pair of hotspot, then this giant radio galaxy might be a double-double radio galaxy. The inner pair of lobes is well aligned
with the outer ones.  However, in most double-double radio galaxies, the inner hotspot is narrower compared to the outer one owing to the
clearing of the path due to the previous ejection. The variation of
intensity along the ridgeline of the giant radio galaxy at 610 MHz is plotted as a
function of distance from the core in Figure~\ref{spatialprofile}. This
plot clearly shows the asymmetry of the giant radio source. The point c
marks the location of the core. Points a and e correspond to seemingly
the first epoch of radio emission. However, there is no hotspot towards
the north and also the emission fades away after the peak of
brightness. There is a difference of 0.2 Mpc in the total lengths of the lobes
from the core to both the sides as is seen in Figure~\ref{spatialprofile}. The brightness peaks corresponding to the 2nd epoch are
also displaced by 0.1 Mpc distance. This difference might be arising due
to the difference in the environments in which they reside. This might indicate that the hotspot advance speed is much lower towards the south
due to a denser environment. The radio power near the FR-I/FR-II divide and also the absence of hotspots towards the northern
side makes the classification of this object as FR-II questionable. This absence could also be an after effect of a disruption in continuous energy supply from the central engine combined with a low-density environment.

% One thing to note is that it is at a comparatively high redshift. The losses due to the inverse-Compton scattering are prominent. The fact that the primary lobe is still emitting (whatever luminosity) points to a higher power compared to the second ejection? The presence of hotspots towards the north is debatable. It seems much more diffuse. 
%However it couldn't be classified as an HYMOR because of the absence the flaring close to the core which is otherwise usually seen in typical FRI type galaxies. %This asymmetry in both the lobes points to the differences in environments they reside in. The distances to the brightness peaks might be different, or it has been already brought to a stop. The reason for 

%Make the flux density versus distance and spectral index vs distance plots.There is some prblem in the plots ...make sure.
%HYMORs

\subsection{Spectral Ageing Analysis}

\begin{figure}
\begin{center}
\begin{tabular}{cc}

\includegraphics[width=7cm]{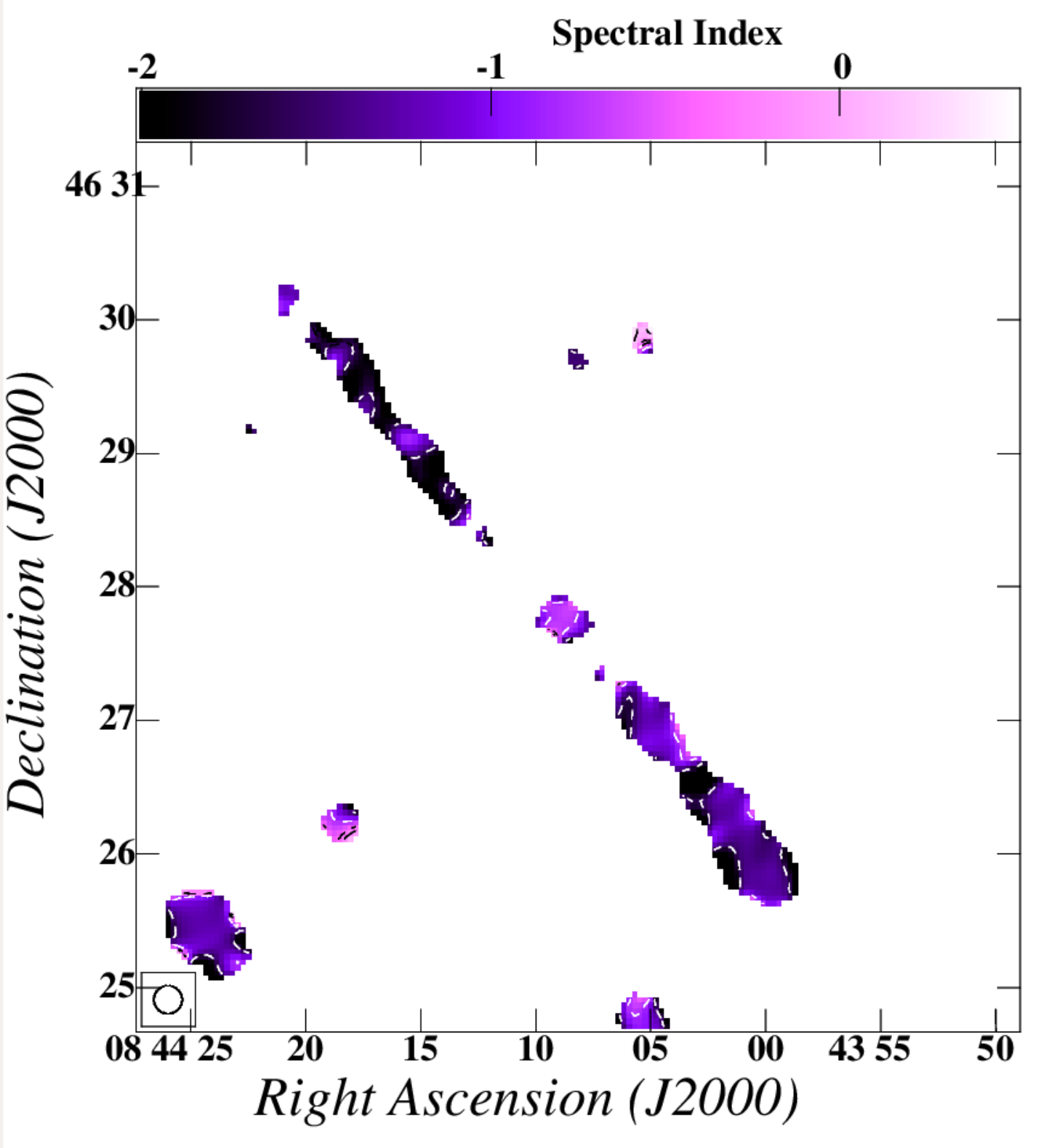}  
\end{tabular}
\caption{325-610 MHz Spectral index map of the giant radio galaxy. }%{\it maybe could have this image in color. }}
    \label{spix}
\end{center}
\end{figure}

The higher energy electrons from a synchrotron radiation emitting population lose their energy faster compared to the low energy ones. As a result, there is a steepening of the energy distribution of the electron population which in turn leads to steepening of the spectrum over time, and the depletion of the high-energy electrons also introduces a break in the radio spectrum. The break frequency shifts to lower frequencies over time. The age of the synchrotron plasma can be estimated by using the value of the spectral index and also the value of the break in the spectrum. Hotspots are thought to provide in-situ acceleration, and it is assumed that the electrons are not accelerated anymore after it leaves the hotspot. Hence in typical FR-II radio galaxies, the spectrum is flatter at hotspots and it then steepens in the lobe towards the core.

Figure~\ref{spix} shows the spectral index map made using images at 325 and 610 MHz. At the position of the hotspot which is suspected to have formed as a result of the second epoch of injection, the spectrum is flatter compared to the regions beyond it. This flattening in spectra is also seen towards the northern lobe as well. The increase in luminosity at these positions as was elaborated in the previous section coupled with the flattening of the spectrum makes it extremely likely that it is indeed a double-double radio galaxy. It is hard to explain the abrupt increase in net energy of electrons in the backflow, without invoking injection of a fresh population of electrons. 

The spectral ageing analysis could act as a powerful tool to probe
whether the object is indeed a double-double radio galaxy. The new
hotspot would be expected to contain a younger population of electrons
due to fresh injection. The difference in spectral age would also give
a rough idea about timescales of episodic emission in such galaxies.

Spectral age is defined as the time which has elapsed since the electrons were last accelerated. Hence, the hotspots are expected to have the least spectral age. The net radiative losses from the hotspots and lobes are attributed to synchrotron and inverse-Compton loss mechanism. The adiabatic losses are neglected. Though it doesn't change the steepening of the spectrum, it shifts the break frequency to lower values. This may lead to the over estimation of the spectral age.

\begin{equation}
\tau_{spec}=\frac{B^{1/2}}{B^2+B_{ic}^{2}}{(\nu_{br}(1+z))}^{-1/2}
\label{eq:1}
\end{equation} 
 
where $B_{ic}=0.318(1+z)^{2}$ is the equivalent magnetic field of cosmic
microwave background radiation at redshift z, B is the magnetic field
strength of the lobes, $\nu_{br}$ is the break frequency in GHz. Both
magnetic fields are expressed in units of nT \citep{2008MNRAS.385.1286J}.
  
As is clear from the equation, one needs to know the magnetic field
at different locations and also the break frequencies to estimate the
age. Equipartition magnetic field was used as a proxy for the physical
magnetic field assuming minimum energy conditions\citep{1975gaun.book..211M}.

\begin{equation}
$$B_{eq}=2.3(aAL/V)^{2/7}$$
\end{equation}

\begin{figure}
\begin{center}
\includegraphics[width=6cm,height=6cm]{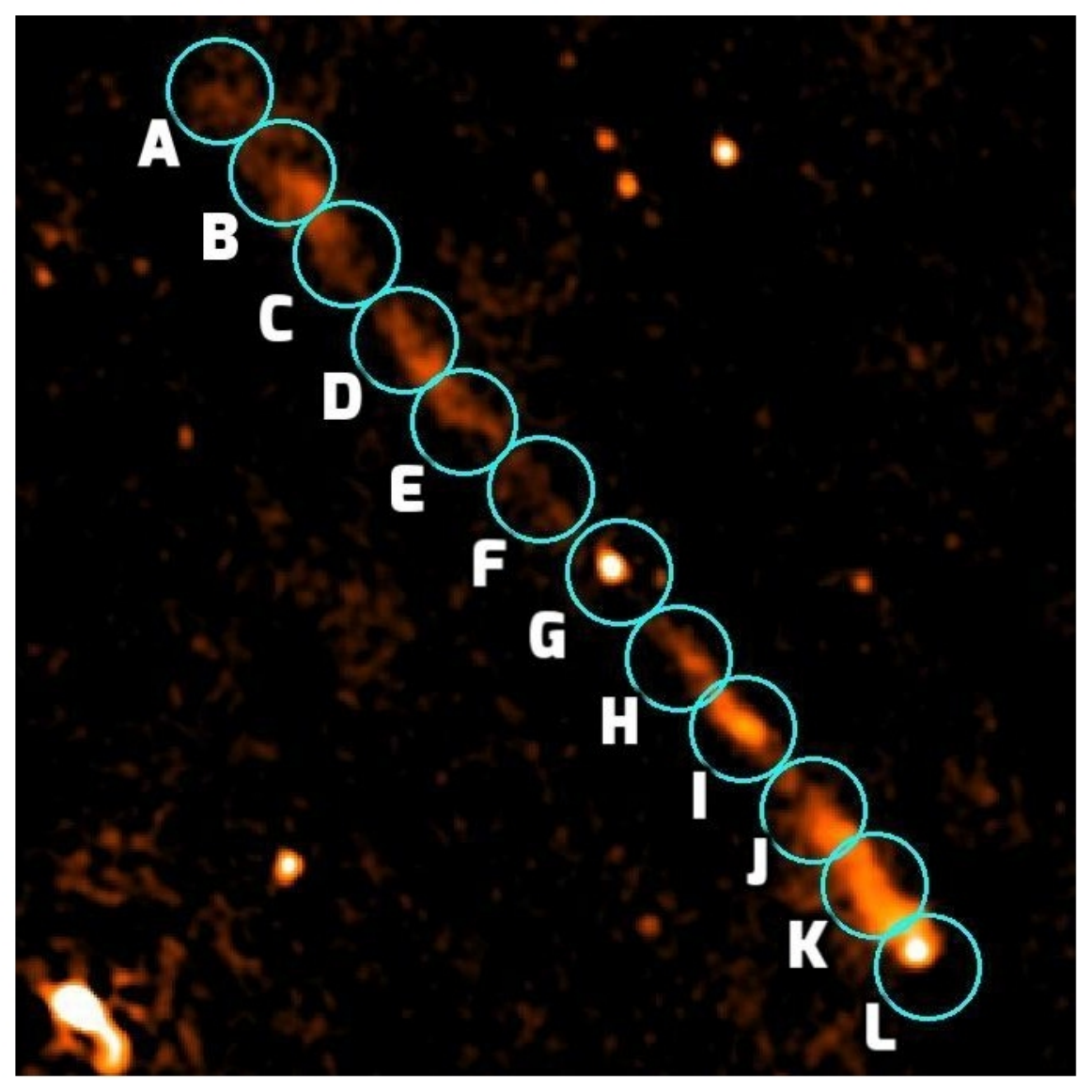}
\\

\caption[Different regions taken to calculate spectral age]
{Different regions taken to calculate spectral age}
\label{magfield}
\end{center}
\end{figure}

Since the giant radio galaxy covers a significant number of resolution
elements, spectral ageing analysis can be done for different regions
separately. The giant radio galaxy was split into 12 circular regions as
shown in Figure~\ref{magfield}.  The flux values were noted for each region
separately except for region G which marks the postion of the core, and the spectral index value for each region was calculated by fitting a power law. Interestingly none of the regions showed any
curvature or deviation from power law at higher frequencies(i.e., L-band),
which means the break is at much higher frequencies. The spectral index
values, luminosities and volume of each region was used to calculate the
equipartition magnetic field at all the regions. Cylindrical symmetry
was assumed while calculating the magnetic field. The ratio of a number of
electrons to protons was chosen to be 1. The magnetic field at different
regions is listed out in the Table~\ref{tab32} except for region G which marks the position of the core. 
\\

\begin{table}

\caption{Values of Equipartition Magnetic field estimated at different regions}
\begin{tabular}{ p{3cm} p{4cm}}

\\
\hline
 Region & Magnetic Field($\mu$G)  \\ 
\hline
 A & 1.07$\pm$0.22\\  
 B & 1.08$\pm$0.21\\  
 C & 1.00$\pm$0.29\\  
 D & 1.24$\pm$0.22\\
 E & 0.80$\pm$0.10\\
 F & 0.57$\pm$0.13\\

 %G & 0.86$\pm$0.08\\
 H & 0.74$\pm$0.08\\
 I & 1.02$\pm$0.04\\
 J & 0.99$\pm$0.09\\
 K & 1.30$\pm$0.20\\
 L & 1.16$\pm$0.15\\
  
\\

 \hline

\end{tabular}\\
\label{tab32}
\end{table} 
The spectra followed a power law from 325 MHz to 1.4 GHz and since there was no evidence for any break up to 1.4 GHz, we
didn't fit the spectrum with standard models like the
Jaffe$–$Perola (JP; \cite{1973A&A....26..423J}) and the Kardashev$–$Pacholczyk
(KP; \cite{1962SvA.....6..317K}; \cite{1970ranp.book.....P}) models.  Instead, 1.4 GHz was chosen as the lower limit on the break frequency,
which puts an upper limit on the spectral ages. 

This upper limit was calculated using the equation~(\ref{eq:1}).

The upper limit on ages ranges from 13 Myr $~$ 20 Myr. \cite{2008MNRAS.385.1286J} finds spectral ages in the range of 6 to 36 Myr and a median spectral age of 22 Myr for a sample of eight giant radio galaxies. Though an upper limit, the spectral age of the giant radio galaxy suggests that it fits in with the general population of giant radio galaxies.
%TILL HERE
%Since we were not able to get an exact

%{\it referee may ask why we did not obtain high-frequency VLA observations to get the break frequency. Hope you have a good answer to this question.}

\subsection{Core Spectrum}

\begin{figure}
\begin{center}
\includegraphics[width=8cm,height=6.5cm]{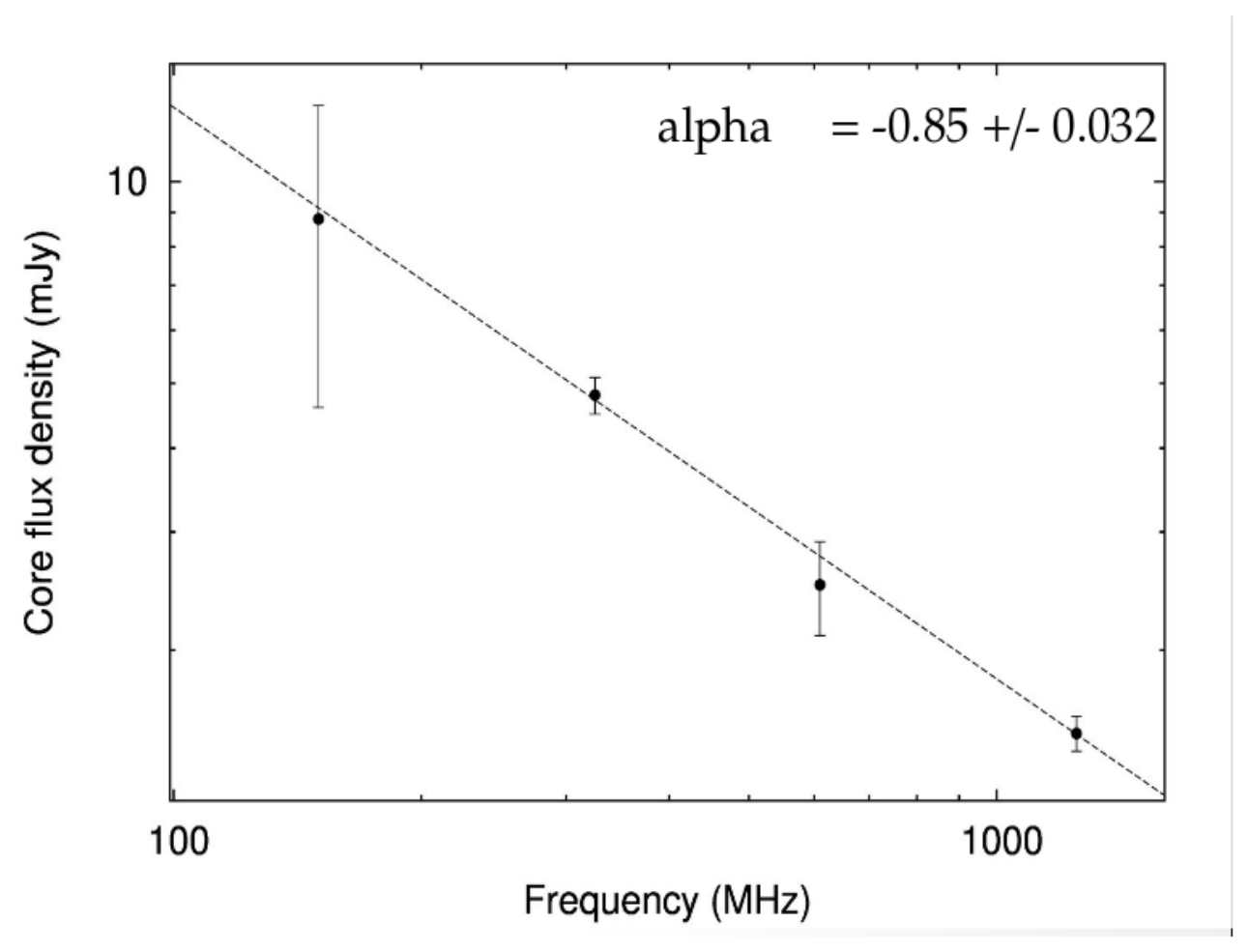}
\\

\caption[Spectrum of the core]
{Spectrum of the core}
\label{corespect}
\end{center}
\end{figure}

The core is unresolved in all the four frequencies from 150 MHz to 1.4 GHz. The core spectrum is unusually steep($\alpha =
-0.85$) for a compact object. There is no turnover even at 150 MHz. In general, for a compact core of AGN, one would expect it to have a flatter spectrum and show turnover or evidence of turnover at lower frequencies due to synchrotron self absorption. The highest resolution image corresponds to 3 arcsec. This angular width corresponds to about 20 kpc. This suggests the presence of yet another pair of lobes near the core. High frequency observations with sub-arcsec resolution could resolve another pair of lobe if any. The core spectrum is shown in Figure~\ref{corespect} and the core flux densities are listed in Table~\ref{tab2}. 
\begin{table}
\caption{Values of the flux density of the core at different frequencies}
\label{tab2}
\begin{tabular}{ p{3cm} p{2cm} p{2cm}}
\\
\hline
 Frequency(MHz) & Core Flux density(mJy) &  Error \\ 
\hline
\\150 & 9.2  & $\pm$4.2
\\325 & 4.8  & $\pm$0.3
\\610 & 2.5 &  $\pm$0.4
\\1250& 1.5  & $\pm$1.0

\\
 \hline

\end{tabular}\\
\end{table}
\section{Discussion and Conclusions}
We have discovered a giant radio galaxy in the field of Lynx.
The giant radio galaxy was detected at all the three observed frequencies including 325, 610 and 1250 MHz. The redshift was estimated using optical observations taken by IUCAA Girawali Observatory and was later confirmed with SDSS DR12 data release. The angular size of 5.5 arcmin translates to a linear size of 2.2 Mpc at the redshift 0.57.
The total flux density of the giant radio source at 325 MHz is  82.3 mJy which corresponds to a specific luminosity of 1.07x10$^{33}$ \,ergs\,s$^{-1}$\,Hz$^{-1}$. This giant radio source is linear in structure and shows no evidence for distortions in the bridges.  The contour maps show evidence for edge brightening in the lobe towards the south, while the hot spot is not very clearly identifiable in the northern lobe. This edge brightening almost halfway the full length of the giant radio galaxy might be due to a difference in environments. The IGM might be relatively less dense to the northern side. Given that the lobes are separated by distances as large as 2 Mpc, it is possible that the environments are considerably different. The equipartition magnetic field in the lobes of the giant radio galaxy was estimated to have an average value of 1.00$\mu$G. Using the estimated equipartition magnetic field at different regions and a lower limit on the break frequency at 1400 MHz, an upper limit of spectral age was estimated as 20 Myr. The giant radio galaxy shows some evidence for episodic emission. The steep-spectrum of the core also suggests the presence of yet another pair of lobes unresolved within the core and this giant radio galaxy turns out to be a candidate triple-double radio galaxy.

\section*{Acknowledgements}

We thank the anonymous referee for her or his comments and suggestions which helped in improving the manuscript.
We thank the staff of the GMRT that made these observations possible. GMRT is run by the National Centre for Radio Astrophysics of the Tata Institute of Fundamental Research.
We wish to acknowledge the IUCAA/IGO staff for their support during our observations.

Funding for the Sloan Digital Sky Survey IV has been provided by
the Alfred P. Sloan Foundation, the U.S. Department of Energy Office of
Science, and the Participating Institutions. SDSS-IV acknowledges
support and resources from the Center for High-Performance Computing at
the University of Utah. The SDSS web site is www.sdss.org.

SDSS-IV is managed by the Astrophysical Research Consortium for the 
Participating Institutions of the SDSS Collaboration including the 
Brazilian Participation Group, the Carnegie Institution for Science, 
Carnegie Mellon University, the Chilean Participation Group, the French Participation Group, Harvard-Smithsonian Center for Astrophysics, 
Instituto de Astrof\'isica de Canarias, The Johns Hopkins University, 
Kavli Institute for the Physics and Mathematics of the Universe (IPMU) / 
University of Tokyo, Lawrence Berkeley National Laboratory, 
Leibniz Institut f\"ur Astrophysik Potsdam (AIP),  
Max-Planck-Institut f\"ur Astronomie (MPIA Heidelberg), 
Max-Planck-Institut f\"ur Astrophysik (MPA Garching), 
Max-Planck-Institut f\"ur Extraterrestrische Physik (MPE), 
National Astronomical Observatories of China, New Mexico State University, 
New York University, University of Notre Dame, 
Observat\'ario Nacional / MCTI, The Ohio State University, 
Pennsylvania State University, Shanghai Astronomical Observatory, 
United Kingdom Participation Group,
Universidad Nacional Aut\'onoma de M\'exico, University of Arizona, 
University of Colorado Boulder, University of Oxford, University of Portsmouth, 
University of Utah, University of Virginia, University of Washington, University of Wisconsin, 
Vanderbilt University, and Yale University.

%{\it Full SDSS acknowledgments need to be added}

%%%%%%%%%%%%%%%%%%%%%%%%%%%%%%%%%%%%%%%%%%%%%%%%%%

%%%%%%%%%%%%%%%%%%%% REFERENCES %%%%%%%%%%%%%%%%%%

% The best way to enter references is to use BibTeX:

%\bibliographystyle{mnras}
%\bibliography{example} % if your bibtex file is called example.bib

% Alternatively you could enter them by hand, like this:
% This method is tedious and prone to error if you have lots of references

%%%%%%%%%%%%%%%%%%%%%%%%%%%%%%%%%%%%%%%%%%%%%%%%%%

%%%%%%%%%%%%%%%%% APPENDICES %%%%%%%%%%%%%%%%%%%%%

%%%%%%%%%%%%%%%%%%%%%%%%%%%%%%%%%%%%%%%%%%%%%%%%%

% Don't change these lines
\bsp % typesetting comment
\label{lastpage}
\end{document}